# Resonant enhancement of Raman scattering in metamaterials with hybrid electromagnetic and plasmonic resonances

**Sriram Guddala[1,\*], D. Narayana Rao[2] and S. Anantha Ramakrishna[1]**

[1] *Department of Physics, Indian Institute of Technology Kanpur, Kanpur-208016, India*
[2] *School of Physics, University of Hyderabad, Hyderabad-500046, India*
[\*]*sguddala@iitk.ac.in*

**Abstract:** A tri-layer metamaterial perfect absorber of light, consisting of (Al/ZnS/Al) films with the top aluminium layer patterned as an array of circular disk nanoantennas, is investigated for resonantly enhancing Raman scattering from C-60 fullerene molecules deposited on the metamaterial. The metamaterial is designed to have resonant bands due to plasmonic and electromagnetic resonances at the Raman pump frequency (725 nm) as well as Stokes emission bands. The Raman scattering from C60 on the metamaterial with resonantly matched bands is measured to be enhanced by an order of magnitude more than from C60 on metamaterials with off-resonant absorption bands peaked at 1090 nm. The Raman pump is significantly enhanced due to the resonance with a propagating surface plasmon band, while the highly impedance matched electromagnetic resonance is expected to couple out the Raman emission efficiently. The nature and hybridization of the plasmonic and electromagnetic resonances to form compound resonances are investigated by numerical simulations.

## 1. Introduction

There has been growing interest in the development of plasmonic nanostructures with intense localized near fields at nano or micro scales for novel applications in micro lasers[1,2], sensing[3], and energy harvesting etc.[4]. Much attention has been devoted to the understanding of plasmonic resonances in various kinds of plasmonic structures[5–7]. Apart from the localized surface plasmons on nanostructures and delocalized surface plasmons supported at metal-dielectric interfaces, the coupling between these two resonances have attracted considerable interest across a wide variety of systems, ranging from coupled microresonators[8] to resonance shifts, hybridized modes and Fano line shapes[9–13], etc.

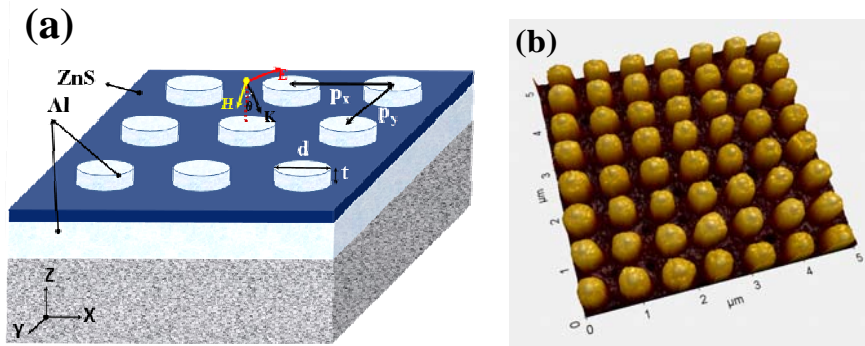

Fig. 1: (a) Schematic unit cell of the tri-layer (Al/ZnS/Al) MPA with spacer ZnS layer thickness of 50 nm and bottom Al layer thickness of 100 nm. (b) AFM image of the top layer nanoantenna Al disks with each disk diameter of 430 (±5) nm and 30 nm thickness.

Recently, metal-dielectric tri-layer systems with a periodic array of metallic nanoparticles as one layer separated from a thin/thick metallic film by a thin insulating layer have attracted great attention due to its special character as a metamaterial perfect optical absorber (MPA) (artificial black body) by optimal impedance matching with the surrounding medium[14]. These novel materials are based on the excitation of electromagnetic resonances or plasmonic resonances in the structure, and are tested for applications in thermal imaging[15], thermo-photovoltaics[4] and sensing[3], etc. This structured metal-dielectric-metal layer system can also act as a nonlinear absorber when interacting resonantly with a high power laser[16]. Farhang et.al,[6] reported compound plasmon resonances due to coupling in a metallic nanograting placed close to a thin gold film. Wang and co-workers reported numerical results on broadband metal-dielectric tri-layer MPA for optical region. A broadband of perfect absorption with 54nm is obtained as a resultant of coupling between multiple plasmonic resonances at the optical frequencies[17]. Chu and co-workers[18] reported enhanced Raman scattering from benzenethiol molecules placed on an array of gold nano disks separated from a thick gold film by a thin dielectric layer due to coupled resonances. Surface enhanced Raman scattering (SERS) has been widely reported from resonant plasmonic nanostructured materials due to localized and enhanced near-fields[19–23]. SERS enhancement factors of greater than $10^6$ were obtained in these plasmoic structures with optimized structural parameters. In these structures, either the Raman pump frequency or Raman band frequencies were superposed with the localized plasmonic resonance frequencies. In this study, a metamaterial perfect optical absorber structure is employed for the surface enhanced Raman scattering studies with the following added advantages than the others. First, the MPA structure provides resonant interaction for both the pump and Raman emission wavelengths, whereby the incident and emitted light can be amplified by great orders of magnitude. Second, the periodicity of the structure can couple the light to free space either by diffraction or by the impedance matching. Third, the resonance tuning by structural design can place the interested molecules in the most intense near-field region.

Here, we used an optimally impedance matched multi-resonant metal-dielectric-metal tri-layer (Fig. 1a) for enhanced Raman scattering from deposited C60 molecules. The system supports electromagnetic and plasmonic resonances at the excitation wavelength as well as the emitted Stokes shifted wavelengths. The origin of the hybrid resonances of the system and their roles for the amplification of the Raman signal are investigated through both experiments and finite element method simulations.

**2. Experimental**

The Al/ZnS/Al system with the schematic unit cell as shown in Fig. 1a consists of 30 nm thick periodic Al disks separated from the 100 nm thick bottom Al film by a 50nm ZnS dielectric layer. The fabrication details can be found in Ref.[16]. The surface morphology was investigated by atomic force microscope (AFM) (Park, XE 70). The Al disk diameter was 430 (±5) nm with a period of 720 (±5) nm (Fig. 1b). The reflection spectrum was measured by using a UV-Vis/NIR spectrophotometer (JASCO V-670) with a beam of (1mm×3mm) incident at an angle of $\theta$ (Fig. 1(a), for $5^0$ to $55^0$) to the sample surface normal. Each reflection spectrum was normalized to the reflection (≈100%) from a thick and smooth Al film. A 3mM toluene solution of C60 molecules was spin coated on the MPA. The Raman scattering of C60 molecules was collected in the back scattering geometry by a confocal micro-Raman spectrometer (Horiba JobinYvon, LabRAM-HR 800) with an excitation wavelength of 785 nm. A 50× objective lens with a numerical aperture of 0.75 focused the excitation beam of 0.7 mW power to ~1 µm² area of the sample. Each Raman spectrum was averaged over 10 seconds' exposure time. A point to point fluctuation of about ±10% in the Raman intensity was found across the area of the samples.

## 3. Results and discussions

The measured reflection spectrum $R(\omega) = 1-A(\omega)-T(\omega)$ of the MPA is shown in Fig. 2(a). The 100 nm thick bottom Al film, which is much more than the skin depth (~3.5 nm at 785 nm), literally makes the transmission zero. The optimized impedance matching of the tri-layer system shows a large drop in the reflected intensity to less than 1% at 850 nm with a large bandwidth presumably due to inhomogeneous broadening of the resonances. This resonance band encompasses the wavelengths of the Raman pump excitation at 785 nm as well as the C60 Stokes wavelengths at 850 nm and will cause the molecules to resonantly interact with both excitation and Stokes light.

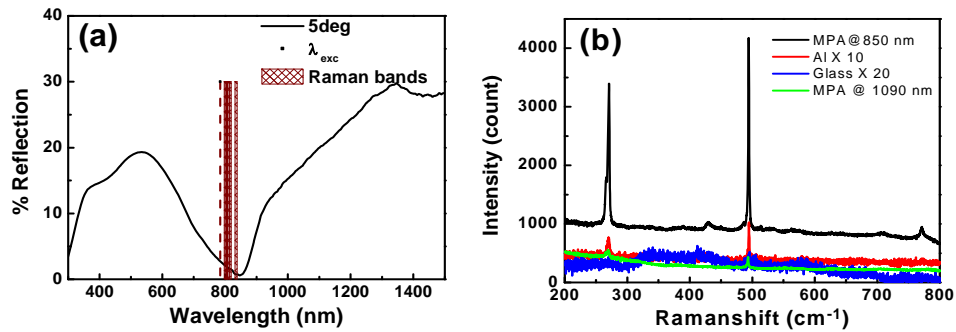

Fig. 2: a) White light reflection spectrum of the MPA measured at 5⁰ angle of incidence. The brown colored dotted line and meshed bars refer to respective spectral positions of Raman excitation wavelength (785 nm) and emitted Stokes wavelengths of C60 molecules. b) Raman scattering signal of 3 mM C60 molecule collected over MPAs with resonances at 850 nm (black) and at 1090 nm (green). The red and blue curves are for the 100 nm thickness Al film and plane glass substrate (blue) for the same analyte concentration and collection configurations.

The enhanced Raman spectrum from the C60 on the MPA in Fig. 2(b) shows four intense bands at 270.7, 428.1, 770.5 cm⁻¹ and 494.1 cm⁻¹, which correspond to respective degenerate $H_g$ and $A_g$ Raman active vibration modes of the C60 molecule [24]. Raman-scattering spectrum from C60 on MPA shows, a strong enhancement of C60 Raman bands[24] for the excitation at 785 nm in comparison to the C60 deposited under identical conditions on bare Al film and glass substrate. The Raman signal from the glass substrate could barely be even measured. The enhancement for the resonant interaction was compared with another MPA with a resonance at 1090 nm for the Al disks larger diameter of (560±5 nm) and period of 880 nm. The wavelengths of excitation and Stokes photons, as shown in Fig. 3b, are off resonant from the strong absorption resonance unlike the earlier MPA. From Fig. 2b, the off-resonant MPA shows an enhancement of only ~ 4 times more with respect to the molecules on a thick Al film. Whereas the molecules on the resonant MPA with a broad resonance at 840 nm shows nearly ~ 41 times larger signal in comparison to the molecules on the thick Al film and approximately 13 times in comparison to the off-resonant MPA. These enhancements show the importance of the resonant interaction with MPA structure and potential of these structures for sensing applications. The resonant enhancement of the Raman signal requires to be investigated for the nature of the resonances involved.

The angle resolved reflection spectrum with the angle of incidence, in Fig. 3a, shows the further broadening of the reflection dip with increase of angle. The full width at half maximum (FWHM) was measured to be varying from 231 nm to 286 nm with the evolution of a new resonance dip. This broadening with angle is associated with another blue shifted resonance band. These two distinguishable bands are only visible in the measurements at higher angles (Fig. 3a), while an asymmetry can still be discerned at lower angles, which indicates the presence of more than one resonance within the broadened absorption band. The enhancement of the Raman signal for the resonant excitation and the experimentally observed angle resolved reflection spectrum broadening and blue shift of the evolved new resonance band can only understood by the MPA structure's simulated spectral response and their near fields.

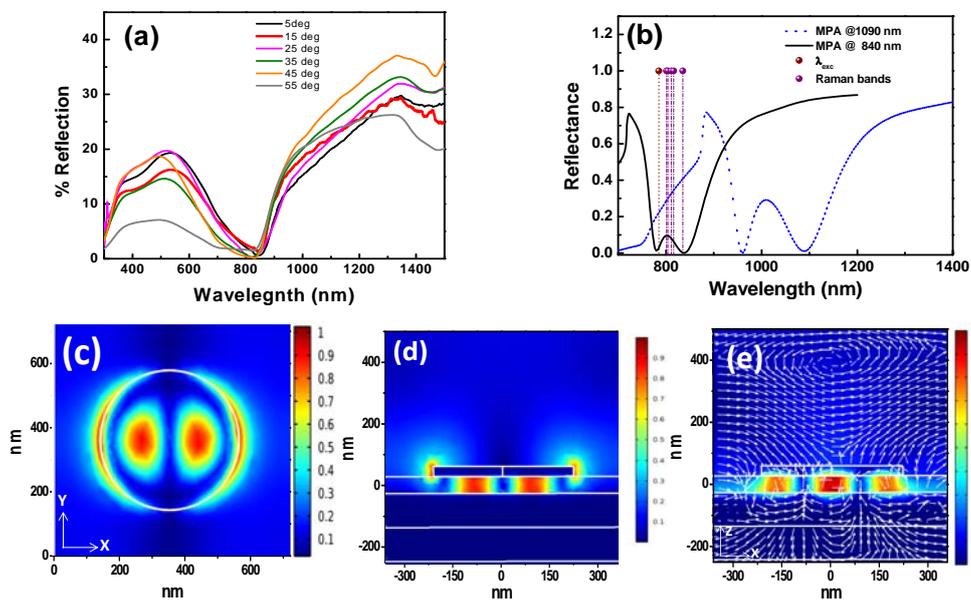

Fig. 3: a) Angle resolved reflection spectrum of MPA with resonance at 850 nm for white light incidence at θ with respect to surface normal (Fig. 1(a)). (b) Simulated spectral response of resonant and non-resonant MPAs; (black) d = 430 nm and period p = 720 nm, (blue) d = 560 nm and period p = 880 nm. The brown colored dotted line and bars refer to the respective excitation and Stokes wavelengths spectral positions. (c) Absolute electric field distributions corresponding to the 3$^{rd}$ order magnetic resonance at 840 nm in (c) the top Al disk and (d) spacer ZnS layer (e) Absolute magnetic field distribution in the spacer ZnS layer for the resonance at 840 nm (Media 1). The time averaged displacement current density in the top and bottom Al layers is shown by the white arrows (Media 1).

The linear absorbance of the MPA can be understood from the finite element method simulations using the commercial COMSOL® Multiphysics. Plane electromagnetic wave at normal incidence to the Al disk array was considered. Periodic boundary conditions were applied along the X and Y directions respectively for the incident TE and TM polarizations. The simulated reflectance spectrum $[R(\omega)=1-A(\omega)-T(\omega)]$ is obtained from the S-parameters of the simulations, where the refractive index (real and imaginary parts) of Al was taken as n = 1.584, k = 10.331[25]. The refractive index of ZnS was taken as n = 2.75 and non-dispersive within the bandwidth considered[26]. The simulated reflectance spectrum is shown in Fig. 3(b) (black line). An absorption band at 850 nm along with another resonance band centered at 780 nm is present for both TE and TM polarizations. For angle dependent simulations (not shown here) the resonance at 780 nm was found to blue shift with respect to the non-dispersive 850 nm resonance. This explains the broad absorption maximum (~98%) noted in the samples at 860 nm. Moreover, the angular dispersion of the resonance at 780 nm with respect to 850 nm resonance confirms the experimental observations of the broadening and then the splitting of the broadband into two at higher angles. Similar two resonant bands are also observed for the MPA resonant at 1090 nm as shown in Fig. 3(b).

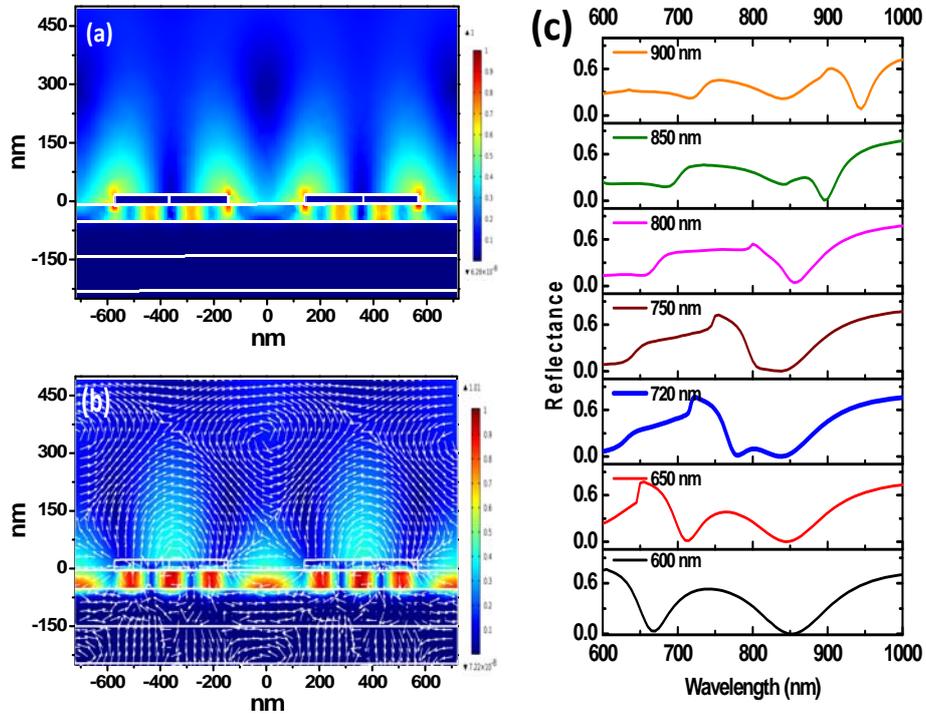

Fig. 4: a) Absolute electric field distribution corresponding to the hybrid resonance at 780 nm. (b) Absolute magnetic field distribution in the spacer ZnS layer for the hybrid resonance (Media 2). The linked media file (Media 2) refers to the time averaged magnetic field and displacement current density distributions (white arrows). (c) Simulated reflectance spectrum of the MPA with lattice period variation from 600 nm to 900 nm at fixed disk diameter of 430 nm.

The electric fields excited at the top Al disk (unit cell shown in Fig. 3(a)), corresponding to the resonance at 840nm, indicates a third order cavity-like mode supported by an antenna [16] with an optical length of $m\lambda/2 = n_{Al}.(2r)$ for ($m = 3$). Now, the third order mode induces image-charges in the bottom Al layer (ground plane) separated by the middle ZnS layer. The resultant anti-parallel currents excited in the top and bottom metallic layers (white arrows in Fig. 3(b), Media 1) form a magnetic dipolar of multipolar distribution that can resonantly interact with the magnetic field of incident radiation. The simultaneous excitation of the electric and the magnetic resonances results in an optimal impedance matching, and a high absorption band peaked at 840 nm results. The third order resonance excited with localized electric and magnetic field distributions in the Al disk and the spacer dielectric layer can be seen in Fig. 3(d) and 3(e) respectively. Owing to the localized character of fields, this optical resonance is non-dispersive for both angular and lattice period variation (shown in Fig. 4(c)). The first order dipole resonance of the MPA is present at 4.7 μm (not shown here). The electromagnetic fields for the two resonances of the MPA resonant near 1090 nm are also similar in nature[16].

The distributions of the electric field at the resonance peak of 780 nm in Fig. 4(a) shows intense scattering fields like an antenna at the top Al disk edges along with a propagating surface plasmon polariton at the bottom Al film-ZnS layer interface. The time averaged displacement current density distribution (Media 2) for the resonance at 780 nm shows propagating surface currents forming along the bottom Al film and ZnS interface between the two disks. The periodic top Al disk array provides an additional grating momentum $\Gamma = (2\pi/p)\sqrt{(m^2+n^2)}$ where $p$ is period, $m$ and $n$ are integers, which couples the incident light at 780 nm into a propagating surface plasmon polariton (PSP) at ZnS and Al film interface. For a given diffraction order of $(m, n) = (1, 0)$, the surface plasmon resonance on the bottom Al-ZnS film interface is found to be close to 780 nm determined from the relation

$$\lambda_{res} = \frac{p}{\sqrt{m^2 + n^2}} \left[ \sqrt{\frac{\varepsilon_{ZnS} \cdot \varepsilon_{Al}}{\varepsilon_{ZnS} + \varepsilon_{Al}}} \right] \qquad (1)$$

where $\varepsilon_{ZnS}$ and $\varepsilon_{Al}$ are the corresponding dielectric permittivities.

The coupling between the localized electromagnetic (LEM) resonance and PSP resonance is provided by scattering from the Al nanoparticle. The fields of the resonance at 840 nm have more localized nature (Fig. 3(d)) than the propagating fields of the resonance (Fig. 4(a)) at 780 nm. The LEM resonance nicely couples to the radiation with optimal impedance matching (bright mode), whereas the PSP modes are non-radiating (dark mode)[7]. The coupling due to the proximity of the two resonances essentially hybridizes them and each resonance picks up some characteristics of the other. Thus, the mode at 780 nm shows a radiating character due to impedance matching with current loops corresponding to a magnetic resonance as shown in Fig. 4(b). The surface currents distributions noticed from Media.2 (Fig. 4(a)) indicates the strong coupling of the two resonances. To understand further the strong coupling of these resonances, the simulations were performed for different array periods from 600 nm to 900 nm for a fixed disk diameter of 430 nm. The reflection spectra for increasing period show distinct non-dispersive LEM and dispersive PSP resonances (Fig. 4(c)). While, the LSP mode remains at 840 nm, the PSP mode shifts to 855 nm and shows strong hybridization between the two resonances. A broadband of high absorption can be noticed from 720 to 750 nm period due to the overlap of two resonances. The structural parameters effects on the PSP and LEM resonances were investigated by varying the disk size and spacer layer thickness for a fixed array period. The LEM resonance shows a blue shift with increased spacer layer thickness whereas the PSP resonance shows a red shift. It is evident from an equivalent LC resonant circuit resonance relation $\omega=1/\sqrt{LC}$, the increase in spacer thickness results in reduced capacitance and a blue shift of the LEM resonance. But, the hybrid PSP resonance shows a distinct red shift with increase of spacer layer thickness. From Eq. (1), it is evident that the increase in thickness of the dielectric medium (ZnS) will cause red shift of the PSP resonance. On the other hand, for the decreased top Al disk diameter the LEM and hybrid PSP resonances were found to change (blue shift) and remain unchanged respectively. Here we investigate the use of these MPA structures for enhancing the Raman scattering from C60 molecules by exploiting the intense radiating and localized fields associated with the MPA.

The Raman shifts at 270.7 cm$^{-1}$ and 494.1 cm$^{-1}$ are the respective $H_g$ and $A_g$ Raman active vibration modes of the C60 molecules[24]. These modes see [Fig. 2(b)] large enhancement in resonant MPA in comparison to the non-resonant MPA, as well as non-textured glass and thick Al film deposited substrates. This strong enhancement can be understood from the spectral location of the excitation wavelength and the Raman bands on the simulated reflection spectra of MPAs as shown in Fig. 3(b). The Raman scattering enhancement in MPA structures is an indication of stringent localized and emitting near-fields contributions. It is evident that the excitation wavelength 785 nm is in resonance with the hybrid PSP resonance, whereas the Raman modes are in resonance with the LEM resonance at 840 nm. So the excitation wavelength 785 nm can be amplified by the intense fields of the hybrid resonance with PSP nature as shown in Fig. 4(a), whereas the emission C60 Raman bands will be enabled by the impedance matched nature of the LEM resonance at 840 nm as shown in Fig. 3(d). In the case of the non-resonant MPA, it is obvious that such resonance mediated amplifications of the pump laser field are only moderate for the excitation. Similarly the density of photon states for the Raman bands will also be much lesser due to the impedance mismatch. This is because both the pump laser line (785 nm) and the Raman bands (805 to 850 nm) lie well outside the resonance bands centered at 1090 nm (Fig. 3(b)). Hence, the intensities of the Raman bands for the resonant and the non-resonant MPAs are different by more than 13 times under the same experimental conditions. These investigations clearly show significant role of the intense fields generated by hybrid resonances in MPAs for sensing applications.

## 4. Conclusions

In summary, we have demonstrated a double resonance mediated Raman scattering enhancement in a MPA where there is a field enhancement at the pump wavelength and an optimized impedance match for emission at the Raman wavelength. The tri-layered metamaterial with a circular nanoantenna array exhibits a band of hybrid PSP mode to which the pump couples and the Raman emission is enabled by a LEM resonance. The enhanced Raman scattering occurs due to the resonant nature of the process. The resonant and off-resonant interactions show the evidence of the intense field's contribution in the amplification of weak Raman signals. The experimental observations were thoroughly verified through theoretical simulations. The hybridization of the nearly PSP and LEM resonances results in an anti-crossing revealed by the numerical studies. Moreover, the amplification of the surface plasmons by our impedance matched MPA structures shows the potential of the designs for ultra level sensing applications as well as in the development of novel plasmonic lasers. The SERS signal can be enhanced better in the MPA structures by putting the molecules in the close vicinity of the resonant cavities. In particular, if the molecule could be imbedded in the spacer layer where the electromagnetic fields are maximal, that would maximize the Raman emission enhancement in comparison to depositing it on plasmonic or metamaterial surfaces. We believe that these MPA structures with sharp hybrid PSP and LEM resonances and their localized and radiating electric fields can have immense interests in the development of high efficiency sensors, by incorporating the analyte in the MPA designs.

**Acknowledgments**

This work was supported by Department of Science and Technology, India under grant no. DST/SJF/PSA-01/2011-2012. SG acknowledges IIT Kanpur for a fellowship.